\documentclass[12pt]{article}
\usepackage{epsf}
\usepackage{graphicx}
\usepackage{a4}
\usepackage{amsmath}
\usepackage{amssymb}% open math symbols
\usepackage{cite}		% collapse citations
\usepackage{color}

\def\hybrid{\topmargin 0pt
        \oddsidemargin 0pt
        \headheight 0pt \headsep 0pt
        \textwidth 6.25in       % A4 paper
        \textheight 9.5in       % A4 paper
        \marginparwidth .875in
        \parskip 5pt plus 1pt   \jot = 1.5ex}

\catcode`\@=11
\def\marginnote#1{}

\newcount\hour
\newcount\minute
\newtoks\amorpm
\hour=\time\divide\hour by60
\minute=\time{\multiply\hour by60 \global\advance\minute by-\hour}
\edef\standardtime{{\ifnum\hour<12 \global\amorpm={am}%
        \else\global\amorpm={pm}\advance\hour by-12 \fi
        \ifnum\hour=0 \hour=12 \fi
        \number\hour:\ifnum\minute<10 0\fi\number\minute\the\amorpm}}
\edef\militarytime{\number\hour:\ifnum\minute<10 0\fi\number\minute}

\def\draftlabel#1{{\@bsphack\if@filesw {\let\thepage\relax
   \xdef\@gtempa{\write\@auxout{\string
      \newlabel{#1}{{\@currentlabel}{\thepage}}}}}\@gtempa
   \if@nobreak \ifvmode\nobreak\fi\fi\fi\@esphack}
        \gdef\@eqnlabel{#1}}
\def\@eqnlabel{}
\def\@vacuum{}
\def\draftmarginnote#1{\marginpar{\raggedright\scriptsize\tt#1}}

\def\draft{\oddsidemargin -.5truein
        \def\@oddfoot{\sl preliminary draft \hfil
        \rm\thepage\hfil\sl\today\quad\militarytime}
        \let\@evenfoot\@oddfoot \overfullrule 3pt
        \let\label=\draftlabel
        \let\marginnote=\draftmarginnote
   \def\@eqnnum{(\theequation)\rlap{\kern\marginparsep\tt\@eqnlabel}%
\global\let\@eqnlabel\@vacuum}  }

\def\numberbysection{\@addtoreset{equation}{section}
        \def\theequation{\thesection.\arabic{equation}}}

\def\titlepage{\@restonecolfalse\if@twocolumn\@restonecoltrue\onecolumn
     \else \newpage \fi \thispagestyle{empty}\c@page\z@
        \def\thefootnote{\fnsymbol{footnote}}
	\setcounter{page}{0} }
\def\endtitlepage{\if@restonecol\twocolumn \else  \fi
        \def\thefootnote{\arabic{footnote}}
        \setcounter{footnote}{0}}  %\c@footnote\z@ }
\definecolor{c1}{rgb}{1, 0, 0}
\definecolor{c2}{rgb}{0, 1, 0}
\definecolor{c3}{rgb}{0, 0, 1}
\definecolor{c4}{rgb}{1, 0, 1}
\definecolor{c5}{rgb}{0, 1, 1}

\catcode`@=12
\relax

\def\ie{\hbox{\it i.e.}}

\def\beq{\begin{equation}}
\def\eeq{\end{equation}}
\def\bea{\begin{eqnarray}}
\def\eea{\end{eqnarray}}
\def\EQ{\begin{equation}}
\def\EN{\end{equation}}

\relax
\numberbysection
\hybrid
\begin{document}
%\begin{titlepage}
\begin{center}
{\large\bf Connectivities of  Potts  Fortuin-Kasteleyn clusters and time-like Liouville correlator}\\[.3in] 

{\bf M.\ Picco$^{1}$, R.\ Santachiara$^{2}$, J. Viti$^3$ and G. Delfino$^4$}\\
        % {\bf (1)}
$^1$ {\it LPTHE\/}\footnote[5]{Unit\'e mixte de recherche du CNRS 
UMR 7589.}, % \\
        {\it  Universit\'e Pierre et Marie Curie-Paris6\\
              Bo\^{\i}te 126, Tour 13-14, 5 \`eme \'etage, \\
              4 place Jussieu,
              F-75252 Paris CEDEX 05, France, \\
    e-mail: {\tt picco@lpthe.jussieu.fr}. }\\
  $^2$ {\it LPTMS\/}\footnote[6]{Unit\'e mixte de 
             recherche du CNRS UMR 8626.}, % \\
             \it{Universit\'e Paris-Sud\\
             B\^atiment 100\\
          F-91405 Orsay, France, \\
    e-mail: {\tt raoul.santachiara@lptms.u-psud.fr}. }\\
   $^3${\it Laboratoire de Physique Theorique de l'ENS}\\ 
   \it{CNRS \& Ecole Normale Superieure,\\
            24 Rue Lhomond, F-75231 Paris, France,\\
    e-mail: {\tt jacopo.viti@lpt.ens.fr}. }\\
$^4$ \it{SISSA and INFN,\\
                       via Beirut 265,\\
                       34136 Trieste, Italy,\\
              e-mail: {\tt delfino@sissa.it}. }\
   
\end{center}
%\vskip .04in
\centerline{(Dated: \today)}
\vskip .2in
\centerline{\bf ABSTRACT}
\begin{quotation}

\noindent
Recently, two of us argued that the probability that an FK cluster in the $Q$-state Potts model connects three
given points is related to the time-like Liouville three-point correlation
function \cite{DV}. Moreover, they predicted that the FK three-point connectivity 
has a prefactor which unveils the effects of a discrete symmetry, 
reminiscent of the $S_Q$ permutation symmetry of the $Q=2,3,4$ Potts model.
%Their theoretical prediction has been checked in \cite{Ziff} for the case of percolation,
% corresponding to $Q=1$. 
We revisit the derivation of the time-like Liouville correlator \cite{Za} and  show that this
is the the only consistent analytic 
continuation of the minimal model structure constants. We then present 
strong numerical tests of the relation between the time-like Liouville correlator and percolative
properties of the FK clusters for real values of $Q$.  

%(We add reference to Zam paper in the abstract)%

\vskip 0.5cm 
%\noindent
%{PACS numbers: 75.50.Lk, 05.50.+q, 64.60.Fr}
% PACS 1999 classification: http://www.aip.org/pacs/pacs99/pacscheme.html
\end{quotation}
%\end{titlepage}
\section{Introduction}
Two-dimensional random fractals occupy a special
place as one can apply the powerful tools of complex analysis to tackle the often very
complicated problems arising from their study. An extremely interesting subclass of
two-dimensional random fractal sets is given by the conformally invariant ones
\cite{Durev,BaBe}. This is the situation when invariance under rescaling and rotations is
enhanced to invariance under any conformal (i.e. analytic and invertible) mapping.
Two-dimensional Brownian motion, critical percolation or the contour lines of a free
Gaussian field, to mention some of the most important and rich-of-applications
random processes, belong to this subclass. In general, the conformal random fractals
are strictly related to the geometrical properties of two-dimensional critical systems. This is the case for
the statistical model we will consider in this paper, the $Q$-state Potts model. 
For $Q=2,3,4$, this is the simplest and most studied spin model (the $Q=2$ case is the Ising model) which undergoes 
a continuous phase transition separating a ferromagnetic from a paramagnetic phase. 
The Fortuin-Kasteleyn (FK) clusters \cite{FK} and the spin clusters at the critical point 
are examples of conformally invariant fractals. In this paper,
we will focus on the FK clusters while  the spin clusters will be considered in a further paper \cite{DPSV2}.
%The FK clusters, which define more general random cluster models, 
%are considered also for general real values of $Q>0$.

In two dimensions, conformal invariance puts an infinite number of constraints on the
behavior of the systems satisfying this symmetry. Conformal Field Theory (CFT) aims
to construct the possible two-dimensional massless field theories whose correlation
functions satisfy this infinite set of constraints \cite{DiFra}. The CFT approach,
which is a powerful alternative to probabilistic 
approaches such as SLE, is based on the assumption that the probabilities associated to
conformally invariant fractals are given by CFT correlation functions. 
In the last twenty years, an important series of results 
followed from the combined use of CFT and  Coulomb gas methods \cite{Ni,Durev}: 
the fractal dimension of many random paths as well as
different geometric exponents controlling, for instance, the reunion probability of an
ensemble of self-avoiding walks \cite{DuSa1,DuSa2,DuKw} or 
the area distribution of Ising and Potts clusters \cite{CaZi}, have been obtained.

Despite these great successes, the methods used to study conformally invariant
fractals and the comprehension of their hidden mathematical structures remain, in
many respects, unsatisfactory. On the one hand, many of the results found so far are
for quantities related to two-point correlation functions, while the fine structure of CFT
fully manifests itself only at the level of three- and four-point correlation functions. Exceptions
are the results derived from the complementary SLE \cite{Carev,BaBe,We} or boundary CFTs
approaches \cite{Caper,SiKlZi,SKFZ} which are mainly 
based on the use of Fuchsian-type partial
differential equations satisfied by probability functions. On the other
hand, the effects of discrete symmetries which arise in pure \cite{GaCa,PiSa,DuJaSa} and disordered \cite{BeLeMi,JLPSW} models is
not understood: for example, the behavior of spin domain walls is basically unknown.
A better knowledge of the CFTs describing extended
objects will pave the way to the computation of important unknown observables.

The simplest family of CFTs, the so called minimal models \cite{BPZ}, have been shown to describe 
local observables of critical statistical models. For instance, the minimal model structure constants \cite{Do,DoFa}
determine the short distance expansion of the scaling fields associated to spin or energy density.
As we will show later, there is a unique consistent analytic continuation of the minimal model
structure constants. This analytic continuation,  we will refer to it as the time-like Liouville three-point correlator,
 has been introduced and computed in \cite{Za} to study 
 the matter content of the minimal gravity model. The time-like Liouville theory  \cite{HaMaWi} is a CFT
that corresponds, at the classical level, to the analytic continuation of the standard
Liouville theory \cite{Te}. It is natural to expect that time-like three-point Liouville
functions, which provide the simplest conformal invariant three-point functions generalizing minimal model ones, 
may play a role in the geometric properties of critical models. 

Recently, two of us have argued that the probability that an FK cluster connects three
given points is indeed related to this function\cite{DV}. Moreover, they predicted that the FK three-point connectivity 
has a prefactor which unveils the effects of a discrete symmetry, 
reminiscent of the $S_Q$ permutation symmetry of the $Q=2,3,4$ Potts model. 
Their theoretical prediction has been checked in \cite{Ziff} for the case of percolation,
 corresponding to $Q=1$.  In this paper we will study the relation between the time-like Liouville correlator and percolative
properties of the FK clusters for general values of 
$Q$.  We will check numerically that the connectivity properties of FK clusters are indeed related to the time-like Liouville
correlators and to the $S_Q$ symmetry prefactor.

The paper is organized as follows. In the next section we recall the random cluster representation of the Potts model,
its relation to conformal field theory and the prediction of \cite{DV} for the relation between the three-point connectivity
and the time-like Liouville correlator. Section~3 is then devoted to show how the time-like Liouville structure constants
can be obtained as analytic continuation of the minimal model ones in the Coulomb gas framework. In section~4 we briefly
summarize the duality and symmetry arguments used in \cite{DV} to argue how permutational symmetry manifests in the final
result for the three-point connectivity, and recall the specificity of the case $Q=3$. In section~5 we present the results
of numerical simulations and their comparison with the theoretical predictions, before making few conclusive remarks
in section~6.

\section{Connectivities of Potts clusters}

\subsection{FK representation of the Potts model}
\label{clusters}

We consider the Q-state Potts model on a  lattice $\mathcal{L}$. This model is defined by the partition function
\begin{equation}
\label{Potts_pf}
Z_{\text{Potts}}=\sum_{\{s(x)\}} e^{J\sum_{\langle x, y\rangle}\delta_{s(x),s(y)}}, 
\end{equation} 
where $s(x)$ is a spin variable taking states $s=1, \cdots , Q$ on each site $x$ of $\mathcal{L}$. 
The sum in (\ref{Potts_pf}) is restricted to neighbouring sites  $\langle x,y\rangle$ and  
$\delta_{s(x),s(y)}$ is the usual Kronecker symbol.
The partition function (\ref{Potts_pf}) is invariant under global permutations 
$\sigma\in S_Q$, the symmetric group of $Q$ elements, acting on the lattice variables.
The model undergoes a phase transition (continuous for $Q\leq 4$ in two dimensions) for a critical value of $J=J_c$.
In the numerical simulations that we will present here, we consider the square lattice where 
$J_c=\log(1+ \sqrt{Q})$\cite{Wu}. Rewriting the Boltzmann weight in (\ref{Potts_pf}) as $e^{{J\delta_{s(x),s(y)}}}=
\bigl[e^J-1\bigr]\delta_{s(x),s(y)}+1$,
leads to the so-called Fortuin-Kasteleyn graph expansion  \cite{FK}
\begin{equation}
\label{Z_FK}
Z_{\text{Potts}}={e}^{JE}\sum_{\mathcal{G}\subseteq\mathcal{L}}p_{r}^{n_b}(1-p_r)^{E-n_b}Q^{C},
\end{equation}
with $E$ the total number of bonds in $\mathcal{L},$ $n_b$  the number of occupied bonds in graph $\mathcal{G}$ (FK graph)
and $C$ the number of its connected components (FK clusters).  
At integer $Q$, given a Potts configuration, FK graphs are constructed putting bonds with probability $p_{r}=1-e^{-J}$
 between neighbouring spins in the same state. 
The graph representation (\ref{Z_FK}) allows to analytically  continue the Potts model to real positive values of $Q$,
defining the so-called random cluster model.
At the critical point $J=J_c$, $p_{r}=p_{r}^{*}$,  FK clusters percolate and 
their critical properties determine the critical exponents for the Potts phase transition. The fixed point is described 
by a CFT with central charge \cite{DoFa,Delfino}
\begin{equation}
\label{Potts_cr}
c(\beta)=1-6\frac{(1-\beta^2)^2}{\beta^2},
\end{equation}
where the values of $\beta$ belong to the critical branch $\frac{1}{2}\leq\beta^2\leq 1$ 
and are related to $Q$ as
\begin{equation}
\label{Q_c}
\text{arcos}\frac{\sqrt{Q}}{2}=\pi(1-\beta^2) \quad  \text{for}\quad  \frac{1}{2}\leq\beta^2\leq 1.
\end{equation}

\subsection{Connectivities and scaling behavior at the critical point}

A way of characterizing a random cluster model is through its $n$-point connectivities, i.e. the probabilities $P_{n}(x_1,\dots,x_n)$ that the points $x_1,\ldots,x_n$ belong to the same cluster. In this paper, we are focusing in particular on the three-point connectivity at criticality.
In order to introduce the three-point connectivity let us recall  known results concerning the two-point connectivity.
The scaling behavior (i.e. for separations $|x-y|$ much larger than the lattice spacing) of the probability  $P^{FK}_2(x,y)$ that an FK cluster 
visits the points $x$ and $y$
is assumed to be given by the two-point function of a scalar primary field in a CFT
\begin{equation}
\label{2pt_connectivity}
P^{FK}_2(x,y)=\langle\Phi_{\Delta_{FK}}(x)
\Phi_{\Delta_{FK}}(y)\rangle. 
\end{equation}
In the above equation  $\Delta_{FK}$ is the scaling dimension associated to the FK  cluster.
We assume that the field $\Phi_{\Delta_{FK}}$ is normalized as
\begin{equation}
  \langle\Phi_{\Delta_{FK}}(x)\Phi_{\Delta_{FK}}(y)\rangle={1 \over |x-y|^{2\Delta_{FK}}},
\end{equation}
while the equality in (\ref{2pt_connectivity}) has to been understood up to a non-universal (lattice dependent) normalization.
As we said before, the FK clusters contain the critical properties of the ferromagnetic Potts model. Indeed, 
the probability  $P_2^{FK}(x,y)$   is related to the two-point correlation function of the spin field and the dimension of the spin operator fixes the magnetic exponent of the critical Potts point.  

Using the parametrization $\beta=\sqrt{\frac{p}{p+1}}$, the conformal dimension
of the spin operator of the corresponding $Q$-state Potts model, see (\ref{Q_c}),  is given by 
\begin{equation}
\Delta_{FK}=2\Delta_{\frac{p+1}{2},\frac{p+1}{2}},
\end{equation}
where
\begin{equation}
\label{Kac_f}
\Delta_{n,m}=\frac{[pn-(p+1)m)]^2-1}{4p(p+1)}.
\end{equation}
Note that in the formula (\ref{Kac_f}), reminiscent of the Coulomb gas approach,  $n,m$ and $p$ are general real numbers.
For integers $n$, $m$ and $p$, with $1\leq n\leq p$ and $1\leq m \leq  p-1$ (\ref{Kac_f})
gives the dimensions of the degenerate
primary fields of the minimal $\mathcal{M}_p$ model with central charge
\begin{equation}
 c(p)=1-\frac{6}{p(p+1)},\quad p=2,3,\dots
\end{equation}

In a similar way, we can also define a three-point connectivity $P_3^{FK}(x,y,z)$  as the probability 
that the sites $x$, $y$ and $z$ are in the same FK cluster. It is expressed by a general conformal invariant three-point function
\begin{equation}
\label{3pt_conn}
P_3^{FK}(x,y,z)=\langle\Phi_{\Delta_{FK}}(x)\Phi_{\Delta_{FK}}(y)\Phi_{\Delta_{FK}}(z)\rangle,
\end{equation}
which takes the form
\begin{equation}
\label{fact}
P_3^{FK}(x,y,z)=R_{FK}\sqrt{P_2^{FK}(x,y)P_2^{FK} (x,z)P_2^{FK} (y,z)} \; .
\end{equation}
The spatial dependence 
of the two- and three-point function for any CFT is fixed by global conformal invariance. On the other hand,
the constant $R_{FK}$ is a particular case of a general structure constant
\begin{equation}
\label{C_delta}
C_{\Delta_1,\Delta_2,\Delta_3}=\lim_{x_3\rightarrow\infty}|x_3|^{2\Delta_3}
\langle\Phi_{\Delta_{1}}(0)\Phi_{\Delta_{2}}(1)\Phi_{\Delta_{3}}(x_3)\rangle,
\end{equation} 
which depends on the details of the CFT under consideration. This makes the study of 
the three-point connectivity particularly interesting because, besides being an important geometric observable, it is a probe 
for testing in which way the conformal symmetry is realized.

For a CFT with  central charge (\ref{Potts_cr}) and non-degenerate spectrum of conformal dimensions, the structure constants (\ref{C_delta}) have been computed,
  for general values  $\Delta_1$, $ \Delta_2$ and $\Delta_{3}$, in \cite{Za}. They are given by the function $C(\alpha_1,\alpha_2,\alpha_3)$ in (\ref{Zam_structure}), with the charges $\alpha_i$'s related to the dimension $\Delta_i$'s via the relation (\ref{Dim_char}). 
The structure constants  $C(\alpha_1,\alpha_2,\alpha_3)$ are related to the time-like Liouville theory which can be thought as an analytical continuation of Liouville theory\footnote{Liouville theory is defined by the Euclidean action
$$
S_{L}=\int d^2x~(\frac{1}{16\pi}\partial_{\mu}\varphi\partial^{\mu}\varphi+\mu e^{-b\varphi}),
$$
where $\varphi$ is a bosonic field and $b\in\mathbb R$. The structure constants for this theory were obtained in different ways in \cite{DoOt,ZaZa,Te}.
 Time-like Liouville field theory corresponds to the analytic continuation to imaginary values of $b \to -i \beta$. See \cite{Za,KoPe,HaMaWi,Gi,ScSu} about the relation between Liouville and time-like Liouville structure constants.}. 
It was argued in \cite{DV} that
\begin{equation}
\label{Rfk}
R_{FK}=\sqrt{2}\,C(\alpha_{\frac{p+1}{2},\frac{p+1}{2}},\alpha_{\frac{p+1}{2},\frac{p+1}{2}},\alpha_{\frac{p+1}{2},\frac{p+1}{2}})
\end{equation}
 for general  $p\geq 2$. Henceforth, we will refer to the above relation as DV prediction. As discussed in \cite{DV}, the prefactor $\sqrt{2}$ can be related to 
the existence of an additional discrete symmetry, to be understood as a sort of analytic continuation of 
the permutational symmetry $S_{Q}$ of Potts model for $Q=2,3,4$ to general value of $Q$ (i.e. $p$). 
The validity of the above conjecture has been verified numerically \cite{Ziff} 
for $p=2$, which corresponds to the random percolation model ($c=0$). 
In this paper we compute numerically  $R_{FK}$ for continuous values of  $p$. Before presenting the numerical simulations, 
we devote the next section to the derivation of the structure constant entering (\ref{Rfk}) within the Coulomb gas framework.

\section{Structure constants of generalized minimal conformal models from the Coulomb gas}

Al.~Zamolodchikov determined in \cite{Za} the structure constants $C(\alpha_1,\alpha_2,\alpha_3)$ of conformal field theory with central charge $c<1$ and non-degenerate spectrum of conformal dimensions, for real values of $c$ and of the conformal dimensions\footnote{The name 'generalized minimal models' is introduced in \cite{Za} to refer to such conformal field theory.} (parameterized by the $\alpha_i$'s as in (\ref{Dim_char})). This result was obtained in \cite{Za} within the conformal bootstrap approach which, starting from the assumption of the decoupling of the null vectors of the Kac fields $\phi_{12}$ and $\phi_{21}$, ends up in a set of two functional equations with a unique solution. This solution is expected to be the analytic continuation of the minimal model structure constants computed in \cite{Doc}, which correspond to rational values of central charge and conformal dimensions. The proof that the Coulomb gas result of \cite{Doc} can be analytically continued to reproduce Zamolodchikov's formula,
 however, is far from obvious and was left in \cite{Za} as an important  problem. We devote this section to a detailed derivation of this continuation, completing the discussion given in \cite{Gi} and showing in particular the uniqueness, an issue which is essential when dealing with analytic continuations.

\subsection{Coulomb gas representation}

We consider a CFT based on a Virasoro algebra with  central charge $c(\beta)$ given in (\ref{Potts_cr}).
A very useful representation of a CFT with central charge $c(\beta)$ is the Coulomb gas representation, 
which is written in term of a Gaussian field $\phi(x)$  
with  background charge 
\begin{equation}
2\alpha_0=\beta-\frac{1}{\beta}
\end{equation}
placed at infinity. The principal objects in this theory are the vertex operators
\begin{equation} 
V_{\alpha}(x)=e^{i \alpha \phi(x)}
\end{equation}
which transform as Virasoro primary operators with dimension
 \begin{equation}
\label{Dim_char}
\Delta_{\alpha}=\alpha(\alpha-2\alpha_0)\; .
\end{equation}
The correlation between vertex operators can be easily calculated using Wick theorem
\begin{equation}
\langle\prod_{i=1}^{N}V_{\alpha_i}(x_i)\rangle=\delta_{\sum_i \alpha_i, 2\alpha_0} \prod_{i<j}|x_i-x_j|^{4\alpha_i \alpha_j}.
\end{equation}
In the above equation, the delta Kronecker $\delta_{\sum_i \alpha_i, 2\alpha_0}$ ensures the vanishing of  
the correlation  if the charge neutrality 
condition 
\begin{equation}
\sum_{i}\alpha_i=2\alpha_0,
\label{cnwc}
\end{equation}
is not satisfied. In order to compute a general  function $\langle\prod_{i}\Phi_{\Delta_i}\rangle$, one can replace  each primary by   
 one of the two vertex operators $V_{\alpha}$ or $V_{2\alpha_0-\alpha}$ which has the same conformal dimension $\Delta_{\alpha}$ 
 \begin{equation}
 \label{rep_mu}
\Phi_{\Delta_{\alpha}}=N_{\alpha}V_{\alpha} \quad \mbox{or} \quad \Phi_{\Delta_{\alpha}}=N_{2\alpha_0-\alpha}V_{2\alpha_0-\alpha}\; .
\end{equation}
As we will see below, the normalization constants $N_{\alpha}$ are highly non-trivial and they are needed to fix the 
ambiguities coming from the identification of $\Phi_{\Delta_{\alpha}}$ with two different vertex operators.  
These constants are strictly related to the so-called \lq\lq exponential \rq\rq normalization in generalized minimal models, see section C in \cite{Za}.

To compute  more general correlation functions,  one would like, by preserving the conformal invariance of the theory, 
 to write a charge neutrality condition
\begin{equation}
\sum_i \alpha_i
+n \beta-m \beta^{-1}= 2\alpha_0
\end{equation} 
which is less strict 
than the one in  (\ref{cnwc}). This can be done  by inserting   into 
the correlation functions two kind of 
screening operators $ V_{\beta}(x)$ and  $ V_{-1/\beta}(x)$ to be integrated all over the plane. This is equivalent 
to modify the 
Gaussian action by adding interaction terms which do not break the conformal invariance. Using
\begin{align}
&&\langle\prod_{i=1}^{N}V_{\alpha_i}(x_i)\prod_{k=1}^{n}V_{\beta}(t_k) \prod_{j=1}^{m} V_{-1/\beta}(\tau_j)\rangle=\delta_{\sum_i \alpha_i
+n \beta-m \beta^{-1}, 2\alpha_0} \prod_{i<j}|x_i-x_j|^{4\alpha_i \alpha_j} \times \nonumber \\
&& \times \underbrace{\prod_{i,k}|x_i-t_j|^{4\alpha_i \beta}\prod_{i,k}|x_i-\tau_j|^{-4\alpha_i/\beta} \prod_{i<j}^{n}|t_i-t_j|^{4 \beta^2}
\prod_{i<j}^{m}|\tau_i-\tau_j|^{4/\beta^2} 
\prod_{k,j}|t_k-\tau_j|^{-4}}_{\equiv D_{n,m}(\{x,t,\tau\})} \nonumber
\end{align}
one has
\begin{equation}
\langle\prod_{i} \Phi_{\Delta_i}(x_i)\rangle\propto\prod_{i<j}|x_i-x_j|^{4\alpha_i \alpha_j}\underbrace{\int \prod_{j=1}^{n} d^2 t_j 
\int  \prod_{k=1}^{m} d^2 \tau_k D^{\{\alpha_i\}}_{n,m}(\{x,t,\tau\})}_{\equiv I_{n,m}(\{\alpha_i\},\{x_i\})}
\end{equation} 
The integral $I_{n,m}(
\{\alpha\},\{x\})$ are the so-called Dotsenko-Fateev integrals. In the following, the explicit $x$ dependence of the integral $I_{n,m}(
\{\alpha\},\{x\})\to I_{n,m}(\{\alpha\})$ will be dropped when the positions of the operators are fixed.

\noindent
\textbf{\underline{Two-point function}}

\noindent
The two-point correlation function $\langle\Phi_{\Delta}(0)\Phi_{\Delta}(1)\rangle$ 
can be written as a Coulomb gas integral if the charges $\alpha$, $\Delta=\alpha(\alpha-2\alpha_0)$, lives on 
the two dimensional lattice
\begin{equation}
\label{operator_content}
 \alpha=\alpha_{n,m}=\frac{1-n}{2}\beta-\frac{1-m}{2\beta} \quad n=0,1,2.. \quad m=0,1,2,..
\end{equation}
The operators $\Phi_{n,m}\equiv \Phi_{\Delta_{\alpha_{n,m}}}$, with conformal dimension $\Delta_{n,m}\equiv \Delta_{\alpha_{n,m}}$  form the primary operator content of the generalized 
 minimal model $\mathcal{M}_{\beta}$ (i.e. $\beta$ a general real number).
 
 There are two possible representations
\begin{equation}
\langle\Phi_{n,m}(0)\Phi_{n,m}(1)\rangle=N_{\alpha_{n,m}}N_{2\alpha_0-\alpha_{n,m}}\langle V_{\alpha_{n,m}}(0)V_{2\alpha_0-\alpha_{n,m}}(1)\rangle
=N_{\alpha_{n,m}}N_{2\alpha_0-\alpha_{n,m}}I_{0,0}(\alpha,
\alpha)=1
\end{equation}
and 
\begin{equation}
\langle\Phi_{n,m}(0)\Phi_{n,m}(1)\rangle=N_{\alpha_{n,m}}^2 I_{n-1,m-1}(\alpha_{n,m},\alpha_{n,m})=1.
\end{equation}
From the above conditions one has therefore 
\begin{equation}
 N_{\alpha_{n,m}}N_{2\alpha_0-\alpha_{n,m}}=1 \quad N_{\alpha}^{-2}=I_{n-1,m-1}(\alpha_{n,m},\alpha_{n,m}).
\end{equation}

\noindent
\textbf{\underline{Three-point function}}

\noindent
We want to compute the three-point function $C(\alpha_1,\alpha_2,\alpha_3)$ in the case all the charges $\alpha_i$ correspond to the discrete set of points 
\begin{equation}
\label{def_alpha}
 \alpha_i=\alpha_{n_i,1}=\frac{1-n_i}{2}\beta\quad \text{with $n_i\in\mathbb{N}$}.
\end{equation}
 We choose the Coulomb gas representation where the operator inserted at infinity is represented by 
$V_{2\alpha_0-\alpha_3}$. In this case, the neutrality condition reads
\begin{equation}
\label{neutrality_Coulombgas}
n\beta=-\alpha_{n_1,1}-\alpha_{n_2,1}+\alpha_{n_3,1} \Leftrightarrow n=\frac{n_1+n_2-n_3-1}{2}\; .
\end{equation}
One has
\begin{eqnarray}
 C(\alpha_{n_1,1},\alpha_{n_2,1},\alpha_{n_3,1})&=&\lim_{z_3\to \infty} |z_3|^{4\Delta_{n_3,1}}
 \langle \Phi_{n_1,1}(0) \Phi_{n_2,1}(1) \Phi_{n_3,1}(z_3)\rangle \\&=&
\sqrt{\frac{N_{\alpha_{n_1,1}}N_{\alpha_{n_2,1}}}{N_{\alpha_{n_3,1}}}} \langle V_{\alpha_{n_1,1}}(0)V_{\alpha_{n_2,1}}(1)V_{2\alpha_0-\alpha_{n_3,1}}(\infty)\rangle 
\nonumber \\
&=&\sqrt{\frac{I_{n_1-1,0}(\alpha_{n_1,1},\alpha_{n_1,1})I_{n_2-1,0}(\alpha_{n_2,1},\alpha_{n_2,1})}{I_{n_3-1,0}(\alpha_{n_3,1},\alpha_{n_3,1})}}I_{n,0}(\alpha_{n_1,1},\alpha_{n_2,1}) \nonumber. 
\label{const_strutt_ninte}
\end{eqnarray}
It is important to stress that the role of the normalization constants is to symmetrize the constant structure $C(\alpha_1,\alpha_2,\alpha_3)$. In the Coulomb gas approach,
the asymmetry is related to the fact that the vertex $V_{\alpha}$ and $V_{2\alpha_0-\alpha}$ are different operators. We are interested in a theory with no multiplicities and that motivates the  identifications (\ref{rep_mu}).
Up to an inessential constant, the integral can be computed explicitly with the result \cite{Doc}
\begin{equation}
\label{cg_int}
 I_{n,0}(\alpha_{n_1,1}\alpha_{n_2,1})=G(1,n,\beta^2,0)\prod_{i=1}^2G(0,n-1,\beta^2,1+2\alpha_{n_i,1}\beta)G(0,n-1,\beta^2,2\beta(n\beta +\alpha_{n_1,1}+\alpha_{n_2,1})-1),
\end{equation}
where we introduced the function $G(x_1,x_2,a,c)$ as
\begin{equation}
\label{prod_gamma1}
G(x_1,x_2,a,c)=\prod_{j=x_1}^{x_2}\frac{\Gamma(aj+c)}{\Gamma(1-aj-c)}\equiv\prod_{j=x_1}^{x_2}\gamma(aj+c), 
\end{equation}
with $\gamma(x)=\frac{\Gamma(x)}{\Gamma(1-x)}$.

\subsection{Analytic continuation of the Coulomb gas integrals to real number $n$ of screenings}

Let us assume for simplicity all the parameters real. The product of Gamma functions
\begin{equation}
\label{prod_gamma}
f(x_1,x_2,a,c)=\prod_{j=x_1}^{x_2}\Gamma(aj+c)
\end{equation}
can be analytically continued to non-integer $x_1$ and $x_2$. In the domain $ax_1+c>0$ and $a(x_2+1)+c>0$ the following integral representation holds
\begin{align}
\label{analytic_cont}
&f(x_1,x_2,a,c)=\text{exp}\left\{\int_{0}^{\infty}\frac{dt}{t}e^{-t}\left[\frac{a}{2}\bigl[x_2(x_2+1)-x_1(x_1-1)\bigr]+
(c-1)(1-x_1+x_2)\right]\right.+ \nonumber
\\
&\left.+\bigl[\frac{e^{-t(ax_1+c)}-e^{-t(a(x_2+1)+c)}}{1-e^{-at}}-e^{-t}(x_2-x_1+1)\bigr]\frac{1}{1-e^{-t}}\right\}\; .
\end{align}
The form (\ref{analytic_cont}) is easily obtained from the formula
\begin{equation}
\label{log}
\log\Gamma(z)=\int_{0}^{\infty}\frac{dt}{t}\left[(z-1)e^{-t}-\frac{e^{-t}-e^{-zt}}{1-e^{-t}}\right].
\end{equation}
It should also be clear that when $x_1$ and $x_2$ do not belong to the domain of convergence of (\ref{analytic_cont}) 
the very definition (\ref{prod_gamma}) can be used to bring $x_1$  and $x_2$ inside the domain\footnote{It is actually straightforward to implement the analytic continuation numerically.}. Notice that the function
$f(x_1,x_2,a,c)$ is meromorphic in the complex plane of $a$ and therefore no issue of analyticity prevents to
continue it from the domain $Re(a)>0$ to $Re(a)<0$.
The analytic continuation
(\ref{analytic_cont}) is however not unique and the ratio of two possible analytic continuations is in general a function with value one when $x_2-x_1$ is an integer
number. For the function $G(x_1,x_2,a,c)$ one finds
\begin{equation}
\label{pgamma}
G(x_1,x_2,a,c)=\frac{f(x_1,x_2,a,c)}{f(x_1,x_2,-a,1-c)}. 
\end{equation}
To make contact with the solution (\ref{Zam_structure}), we introduce \cite{Barnes, TePo}
the Barnes double Gamma function $\Gamma_2(x|\beta,\beta^{-1})\equiv\Gamma_{\beta}(x)$ defined for real $x>0$ and $\beta\not=0$
through the integral representation\footnote{ In the complex plane of $\beta$,
$\Gamma_{\beta}(x)$ is not defined for $Re(\beta)=0$ \cite{Barnes}.}
\begin{equation}
\label{Teschner}
\log\Gamma_{\beta}(x)=\int_{0}^{\infty}\frac{dt}{t}\left[\frac{e^{-xt}-e^{-Qt/2}}{(1-e^{-\beta t})(1-e^{-t/\beta})}
-\frac{(Q-2x)^2}{8e^t}-\frac{Q-2x}{2t}\right],
\end{equation}
with $Q=\beta+\frac{1}{\beta}$. Using (\ref{log}) it is possible to show\footnote{It is also useful to recall the identity
$$
\log x=\int_{0}^{\infty}\frac{dt}{t}\bigl[ e^{-t}-e^{-xt}\bigr].
$$} the recursive relation
\begin{equation}
\label{Teschner_rec}
\Gamma_{\beta}(x+\beta)=\sqrt{2\pi}\beta^{\beta x-1/2}\Gamma^{-1}(\beta x)\Gamma_{\beta}(x).
\end{equation}
From (\ref{Teschner_rec}) and again using (\ref{log})  we can show that for integer $n$ the relation
\begin{align}
\label{Teschner-f}
\log\Gamma_{\beta}(x+n\beta)-\log\Gamma_{\beta}(x)&=\frac{n}{2}\log 2\pi+n(\beta x-1/2)\log\beta+\frac{n(n-1)}{2}\log\beta^2+
\\
&-\log f(0,n-1,\beta^2,\beta x)\nonumber 
\end{align}
holds at the level of the integral representations (\ref{Teschner}) and (\ref{analytic_cont}). 
We now take $n$ real and observe that the $n$ dependence of the terms containing logarithms
in  the right hand side is analytic;
we therefore conclude
that at any given $\beta$ and $x$ real, the analytic continuation
for the  product of $\Gamma$ functions in (\ref{prod_gamma}) satisfies
\begin{equation}
\label{Teschner_f2}
f(0, n-1,\beta^2,\beta x)=(2\pi)^{\frac{n}{2}}\beta^{n(\beta x-1/2)+\frac{n(n-1)}{2}\beta^2}
\frac{\Gamma_{\beta}(x)}{\Gamma_{\beta}(x+n\beta)},
\end{equation}
for arbitrary real $n$.
Starting from the Barnes double Gamma function we can introduce Zamolodchikov's $\Upsilon_\beta(x)$ function as
\begin{equation}
\label{Zam}
\Upsilon_\beta(x)=\frac{1}{\Gamma_\beta(x)\Gamma_{\beta}(Q-x)},
\end{equation}
which has the integral representation convergent in the domain $0<x<Q$
\begin{equation}
\label{Zam_int}
\log\Upsilon_{\beta}(x)=\int_{0}^{\infty}\frac{dt}{t}\left[\frac{(Q/2-x)^2}{e^t}-\frac{\sinh^2\frac{t}{2}(Q/2-x)}{\sinh\frac{\beta t}{2}
\sinh\frac{t}{2\beta}}\right]
\end{equation}
and satisfies the recursive relation
\begin{equation}
\label{rec_Zam}
\frac{ \Upsilon_{\beta}(x+\beta)}{\Upsilon_{\beta}(x)}=\gamma(\beta x)\beta^{1-2\beta x},
\end{equation}
as it follows from (\ref{Teschner_rec}). Notice also that from its definition $\Upsilon_{\beta}(x)=\Upsilon_\beta(Q-x)$.  
The function $G(0,n-1,\beta^2,\beta x)$,  introduced in (\ref{pgamma}) is related for arbitrary real $n,\beta$ and $x$
to $\Upsilon_{\beta}(x)$ by
\begin{equation}
G(0,n-1,\beta^2,\beta x)=\frac{\Upsilon_\beta(x+n\beta)}{\Upsilon_\beta(x)}\beta^{n(2\beta x-1)}\beta^{\beta^2n(n-1)},
\end{equation}
as a consequence of (\ref{Teschner_f2}). We can now rewrite, recalling the neutrality condition (\ref{neutrality_Coulombgas}), 
a possible analytic continuation of $I_{n,0}(\alpha_{n_i,1},\alpha_{n_2,1})$ (see (\ref{cg_int})) to real $n$ and $n_i$ as
\begin{align}
\label{cg_ac}
&I_{n,0}(\alpha_{n_i,1},\alpha_{n_2,1}) \to I(\alpha_1,\alpha_2,\alpha_3)\\ \nonumber  
&I(\alpha_1,\alpha_2,\alpha_3)=\beta^{-2(\beta-\beta^{-1})\alpha_{12}^3}~\frac{\Upsilon_{\beta}(\beta^{-1}+\alpha_{13}^2)
\Upsilon_{\beta}(\beta^{-1}+\alpha_{23}^1)\Upsilon_{\beta}(2\beta-\beta^{-1}-\alpha_{123})
\Upsilon_{\beta}(\beta-\alpha_{12}^3)}
{\Upsilon_{\beta}(\beta^{-1}+2\alpha_1)\Upsilon_{\beta}(\beta^{-1}+2\alpha_2)\Upsilon_{\beta}(2\beta-\beta^{-1}-2\alpha_3)\Upsilon_{\beta}(\beta)}, 
\end{align}
where we used the notations $\alpha_{ij}^k=\alpha_i+\alpha_j-\alpha_k$, $\alpha_{ijk}=\alpha_i+\alpha_j+\alpha_k$.

\subsection{Uniqueness of Zamolodchikov's analytic continuation}
Using the continuation (\ref{cg_ac}) in (\ref{const_strutt_ninte}), we obtain
\begin{equation}
\label{Zam_structure}
C(\alpha_1,\alpha_2,\alpha_3)=A_{\beta}\frac{\Upsilon_{\beta}(\beta-\alpha_{13}^2)\Upsilon_{\beta}(\beta-\alpha_{23}^1)\Upsilon_{\beta}(\beta-\alpha_{12}^3)\Upsilon_{\beta}(2\beta-\beta^{-1}-\alpha_{123})}
{\Bigl[\prod_{i=1}^3\Upsilon_{\beta}(\beta-2\alpha_i)\Upsilon_{\beta}(2\beta-\beta^{-1}-2\alpha_i)\Bigr]^{1/2}}, 
\end{equation}
with the normalization constant
\begin{equation}
\label{Zam_norm}
A_{\beta}=\beta^{-1-\beta^2+\beta^{-2}}\frac{\bigl[\gamma(\beta^2)\gamma(\beta^{-2}-1)\bigr]^{1/2}}{\Upsilon_{\beta}(\beta)}.
\end{equation}
The formula (\ref{Zam_structure}) was first found in \cite{Za}. As we observed earlier the computation of (\ref{Zam_structure}) can be generalized to imaginary $\beta=ib$ and $\alpha_i=ia_i$,
starting from the Coulomb gas integrals and repeating all the steps. The $\Upsilon_b(x)$ can be still introduced observing that
\begin{equation}
G(0,n-1,-b^2,bx)=\frac{1}{G(0,n-1,b^2,1-bx)}.
\end{equation}

In general one can have
\begin{equation}
\label{gen_an}
\tilde{C}(\alpha_1,\alpha_2,\alpha_3)=C(\alpha_1,\alpha_2,\alpha_3)r(\alpha_1,\alpha_2,\alpha_3)
\end{equation}
where the function $r(\alpha_1,\alpha_2,\alpha_3)=1$ if the $\alpha_i$ belong to the set (\ref{def_alpha}) and satisfy the neutrality condition (\ref{neutrality_Coulombgas}), i.e. if the constants are given by (\ref{const_strutt_ninte}). We now wish to show under which reasonable assumptions the analytic continuation (\ref{Zam_structure}) is unique, or in other words when $r(\alpha_1,\alpha_2,\alpha_3)=1$ for any value of the charges $\alpha_i$.

 We first notice that
the integral $I_{n,0}(\alpha_{n_1,1},\alpha_{n_2,1})$ satisfies a recursive relation in $n$ as it can be easily seen from (\ref{cg_int}). We assume these recursion relations 
to hold also for its possible
analytic continuations. Taking into account that decreasing $n$ by one unity, $n\to n-1$, amounts to replace for example $\alpha_1$ with $\alpha_1+\beta$, we obtain the following functional
relation\footnote{The relation above is equivalent to the first functional equation obtained for
the structure constants $C(\alpha_1,\alpha_2,\alpha_3)$ within the so-called conformal bootstrap approach. The conformal bootstrap method relies
on the conformal invariance of the Coulomb gas action and has been first proposed in \cite{Te} in the context of Liouville field theory
for a proof of the DOZZ formula \cite{DoOt,ZaZa}.}
\begin{align}
\label{func1}
&\frac{\tilde{C}(\alpha_1+\beta,\alpha_2,\alpha_3)}{\tilde{C}(\alpha_1,\alpha_2,\alpha_3)}
=\frac{C(\alpha_1+\beta,\alpha_2,\alpha_3)}{C(\alpha_1,\alpha_2,\alpha_3)}= \\ \nonumber 
& =\frac{\gamma(\beta^2-\beta\alpha^1_{23})\gamma(2-\beta^2+\beta\alpha_{123})}
{\gamma(-\beta\alpha^2_{13})\gamma(-\beta\alpha^3_{12})}
\Bigl[\gamma(-1-2\alpha_1\beta)
\gamma(\beta^2-1-2\alpha_1\beta)\gamma(-\beta^2-2\alpha_1\beta)\gamma(-2\alpha_1\beta)\Bigr]^{1/2}. 
\end{align}
This equation implies that $r(\alpha_1,\alpha_2,\alpha_3)$ is periodic (in all its variables)
with period $\beta$. 
\begin{figure}
\begin{center}
\includegraphics[width=14cm]{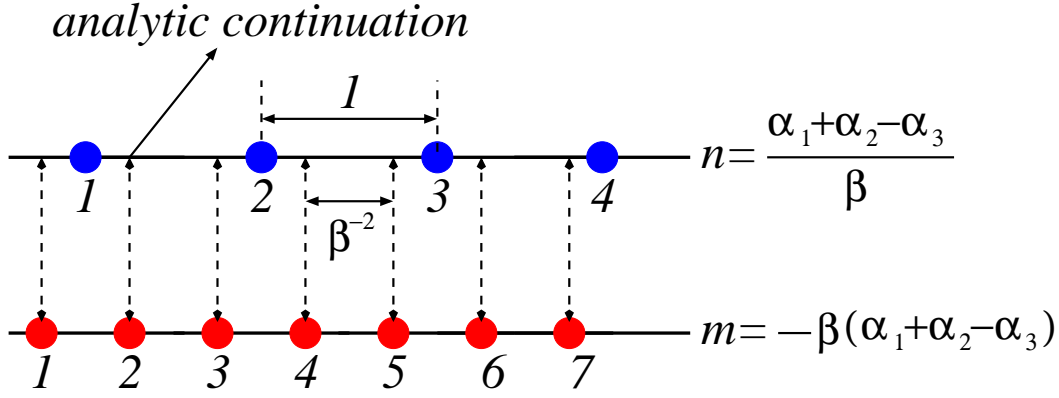}
\end{center}
\caption{
Consider $C(\alpha_1,\alpha_2,\alpha_3)$ given in (\ref{gen_an}) where values
of $\alpha_{2}$ and $\alpha_{3}$ are fixed. The function $C(\alpha_1,\alpha_2,\alpha_3)$ depends then on 
one parameter $n$ represented in the upper axis. The dots represent integer values of $n$
where the neutrality conditions are satisfied and
 the $C(\alpha_1,\alpha_2,\alpha_3)$ are given by Coulomb gas integrals with $n$ screenings of  type $\beta$. In the lower axis, the points correspond to the case where  
 $C(\alpha_1,\alpha_2,\alpha_3)$ is given by Coulomb gas integrals of $m$ screenings of  type
 $-1/\beta$. As $\beta^2$ is assumed irrational, the dashed arrow indicate that  $C(\alpha_1,\alpha_2,\alpha_3)$
 for real values of $n$ should match the Coulomb gas integral with  $m$ screenings of  type $-1/\beta$, as written
 in (\ref{uni_con})}
\label{unicity}
\end{figure}
The next crucial observation is that the generalized minimal model $\mathcal{M}_{\beta}$ must be identified
with  $\mathcal{M}_{-1/\beta}$, which has the same central charge $c(\beta)=c(-1/\beta)$ and
operator content (\ref{operator_content}). We recall that we obtained the structure constants by an analytic
continuation of the Coulomb gas integrals associated to the subset of fields $\Phi_{n,1}$, $n=1,2..$. One has to require
that the analytic continuation gives, at the same time, the structure constants of the fields $\Phi_{1,m}$, $m=1,2..$ 
\begin{align}
\label{uni_con}
\tilde{C}(\alpha_{1+(1-m_{1})/\beta^2,1},\alpha_{1+(1-m_2)/\beta^2,1},\alpha_{1+(1-m_{3})/\beta^2,1})&= \\ \nonumber
&\hspace*{-3cm}=\sqrt{\frac{I_{0,m_1-1}(\alpha_{1,m_1},\alpha_{1,m_1})I_{0,m_2-1}(\alpha_{1,m_2},\alpha_{1,m_2})}{I_{0,m_3-1}(\alpha_{1,m_3},\alpha_{1,m_3})}}I_{0,m}(\alpha_{1,m_1},\alpha_{1,m_2})
\end{align}
where $m_1,m_2,m_3$ and $m$ are positive integers satisfying $m=(m_1+m_2-m_3-1)/2$, see Fig.(\ref{unicity}).
From the above condition, and taking into account the recursion relation for the integrals $I_{0,m}(\alpha_{1,m_1},\alpha_{1,m_2})$ (obtained by replacing $\beta \to-1/\beta$), 
the function $r(\alpha_1,\alpha_2,\alpha_3)$ must also  be 
periodic with period $1/\beta$ in all its arguments.  If $\beta^2$ is irrational such a function is one, since a function of a real variable which has two incommensurable periods is a constant.

\section{Connectivities of FK clusters and duality}
\label{FK_duality}
The FK mapping allows one to express correlation functions of the local Potts spin operator 
\EQ
\sigma_\alpha(x)=Q\delta_{s(x),\alpha}-1\,,\hspace{1cm}\alpha=1,\ldots,Q
\label{spin}
\EN
in terms of connectivities of FK clusters. Consider for  example the one-point function $\langle\sigma_{\alpha}(x)\rangle$ on a simply connected domain
$D\subseteq\mathbb R^2$, on the boundary of which the Potts spins are fixed to have value $\alpha$.
It is not difficult to realize that the FK expansion for $\langle\sigma_{\alpha}(x)\rangle$ only contains configurations in which $x$
is connected to the boundary of $D$ and, in the thermodynamic limit in which $D$ becomes the whole plane, it is related to the percolative order parameter $P^{FK}$ of FK clusters, i.e. the probability that there exists an infinite FK cluster, as
\EQ
\label{one-point}
\langle\sigma_{\alpha}(x)\rangle=(Q-1)P^{FK}.
\EN
Similarly (see e.g. \cite{RandomClu, VaJaSa}) for $J\leq J_c$ the two- and three-point spin correlators are related to the FK connectivities as
\bea
\langle\sigma_{\alpha}(x)\sigma_{\alpha}(y)\rangle &=& (Q-1)\,P_{2}^{FK}(x,y)\,,
\label{G2}\\
\langle\sigma_{\alpha}(x)\sigma_{\alpha}(y)\sigma_{\alpha}(z)\rangle &=& (Q-1)(Q-2)\,P_{3}^{FK}(x,y,z)\,.
\label{G3}
\eea  
Together with duality (see e.g. \cite{RandomClu}), these equations lead to
\bea
P_{2}^{FK}(x,y)&=&\langle\mu_{\alpha\beta}(x)\mu_{\beta\alpha}(y)\rangle\,, 
\label{mu2}\\
P_{3}^{FK}(x,y,z)&=&\langle\mu_{\alpha\beta}(x)\mu_{\beta\gamma}(y)\mu_{\gamma\alpha}(z)\rangle  \,,
\label{mu3}
\eea  
where $\mu_{\alpha\beta}(x)$ are disorder (or kink) fields and their correlators are evaluated at $J^*\geq J_c$.

It was observed in \cite{DV} that the color index structure of the disorder correlators in (\ref{mu2}) and (\ref{mu3}) is in a sense `redundant' because, as a consequence of permutational symmetry, there is only a single two-point correlator and a single three-point correlator, which can be represented as $\langle\mu\mu^\dagger\rangle$ and $\langle\mu\mu\mu\rangle=\langle\mu^\dagger\mu^\dagger\mu^\dagger\rangle$, in terms of a doublet of fields\footnote{It is important to stress that a complete representation of $n$-point correlators of $\mu_{\alpha\beta}$ in terms of $\mu$ and $\mu^\dagger$ is no longer possible for $n>3$, because the number of inequivalent correlators becomes too large. See \cite{RandomClu} for a detailed analysis.} $\mu$, $\mu^\dagger$ satisfying the OPE's $\mu\mu^\dagger=I+\ldots$, $\mu\mu+\mu^\dagger\mu^\dagger=C_\mu(\mu+\mu^\dagger)+\ldots$, where we omit the coordinate dependence; these OPE's express the two-channel structure of the OPE $\mu_{\alpha\beta}\mu_{\beta\gamma}=\delta_{\
alpha\gamma}I+(1-\delta_{\alpha\gamma})C_\mu\,\mu_{\alpha\gamma}+\ldots$, which alone accounts for the two- and three-point correlators. Defining now $\phi=(\mu+\mu^\dagger)/\sqrt{2}$, the two- and three point connectivities become $P_2=\langle\phi\phi\rangle$ and $P_3=\sqrt{2}\langle\phi\phi\phi\rangle$. Substitution into (\ref{fact}) then gives $R_{FK}=C_\mu=\sqrt{2}C_{\phi\phi\phi}$, with $C_{\phi\phi\phi}$ the structure constant of the field $\phi$, which by construction has the conformal dimension of the Potts spin field. It was further argued in \cite{DV} that $C_{\phi\phi\phi}$ should be computable within a CFT with the Potts central charge but without internal symmetry (and then by the Zamolodchikov's formula), because the replacements $\mu_{\alpha\beta}$ $\to$ $\mu,\mu^\dagger$ $\to$ $\phi$ have factorized the color dependence into the $\sqrt{2}$ prefactor.
Before turning to the numerical verification of (\ref{Rfk}) in the next section, let us recall why it certainly holds at $Q=3$.

For $Q=3$ the critical point is described by a particular extended CFT, 
namely the $WA_2(3,4)$ theory.  This is a minimal model of a $WA_2$ current algebra  
which realizes the conformal as well as the permutational $S_3$ symmetry \cite{FaLy}. 
This is important because in the  $WA_2(3,4)$ theory the fields with the same conformal 
dimension but carrying a different $S_Q$ charge are distinguished.
Starting from the definition (\ref{spin}), consider the dual spin variable\footnote{The
dual spin variables are  Potts variables in the dual lattice $\mathcal L^*$.} $\tilde{s}(x)$ and the dual order parameter
$\tilde{\sigma}_{\alpha}(x)=Q\delta_{\tilde{s(x)},\alpha}-1$. Disorder fields are defined by  
\begin{equation}
\label{spin_3}
\mu(x)=e^{-\frac{\pi}{3} i}\tilde{\sigma}_{1}(x)-e^{\frac{\pi}{3} i}\tilde{\sigma}_{2}(x) \quad
\mu^{\dagger}(x)=e^{\frac{\pi}{3} i}
\tilde{\sigma}_{1}(x)-e^{-\frac{\pi}{3} i}\tilde{\sigma}_{2}(x), 
\end{equation}
 and they have  $Z_3$ charges $1$ and $-1$. In the continuum limit,
these fields are associated to the $WA_2$ highest weight representations   $\Phi_{(12)|(11)}$ and $\Phi_{(21)|(11)}$ respectively 
\footnote{The notation $\Phi_{(n m)|(n' m')}$ for the $WA_2$ theories comes from its two-component Coulomb gas representation.}
\begin{equation}
 \mu(x)\to \Phi_{(12)|(11)}(x) \quad \mu^{\dagger}(x) \to \Phi_{(21)|(11)}(x)\; .
\end{equation}
The structure constants of the $WA_2(3,4)$ theory take into account the $S_3$ multiplicities
which determines the correlation functions of the spin fields (\ref{spin_3}). It follows
\begin{eqnarray}
\label{FK_q3}
\left.R_{FK}\right |_{Q=3}&=&\lim_{x_3\rightarrow\infty} x_3^{2\Delta_{FK}}\langle \Phi_{(12)|(11)}(0)\Phi_{(12)|(11)}(1)\Phi_{(12)|(11)}(x_3)\rangle \nonumber \\ 
&=&\lim_{x_3\rightarrow\infty} x_3^{2\Delta_{FK}}\langle \Phi_{(21)|(11)}(0)\Phi_{(21)|(11)}(1)\Phi_{(21)|(11)}(x_3)\rangle \nonumber \\ 
&=&\sqrt{\gamma(1/5)[\gamma(3/5)]^3} \; .
\end{eqnarray}
The above value can be easily related to 
the structure constant $C(\alpha_{3,3},\alpha_{3,3},\alpha_{3,3})$ of the minimal model $M_5$.  One observes indeed that the field $\Phi_{3,3}$
 corresponds to the following real combination
\begin{equation}
\label{neutral}
\Phi_{3,3}=\frac{1}{\sqrt{2}} \left( \Phi_{(21)|(11)}+\Phi_{(12)|(11)}\right),
\end{equation}
which simply implies
\begin{equation}
\label{structure_constant}
\left.R_{FK}\right|_{Q=3}=
\langle \Phi_{(12)|(11)}(0)\Phi_{(12)|(11)}(1)\Phi_{(12)|(11)}(\infty)\rangle=\sqrt{2} C(\alpha_{3,3},\alpha_{3,3},\alpha_{3,3}).
\end{equation}

\section{Numerical simulations}
We will now present results of numerical simulations for  FK clusters.  
Our aim is to test DV prediction (\ref{Rfk}). 
For all the simulations presented in this paper, 
we consider a square lattice of linear size $L$ and with periodic boundary condition in both directions. 
For $Q > 1$, we employ a cluster algorithm to equilibrate the system. For integer $Q$'s, it is the usual Wolff algorithm 
\cite{Wo}, while for non integer $Q$'s, we employ the  Chayes-Machta algorithm 
\cite{CM,DGMOPS} which is an extension of the Swendsen-Wang algorithm. For each value of $Q$,  
 we perform measurements over $10^6$ independent configurations  for $L \leq 2048$, $2.10^5$ for $L=4096$ and $10^5$ for $L=8192$. 
The computation of the three-point correlation function is done by considering, for each point $(x,y)$ 
on the lattice, the correlation  $P^3_{FK}((x,y),(x + \Delta,y),
(x,y+\Delta))$ as a function of $\Delta$.
In the following, we will first present our numerical results for the $Q$-state Potts model with $Q$ integer. 
We then show additional results for the cases where $Q$ is non integer. All our numerical results are 
collected in Tab.~\ref{Table1} where we also show DV predictions.

\noindent 
{\bf \underline{Percolation, $Q=1$}}

\noindent 
\begin{figure}
\begin{center}
\epsfxsize=400pt\epsfysize=300pt{\epsffile{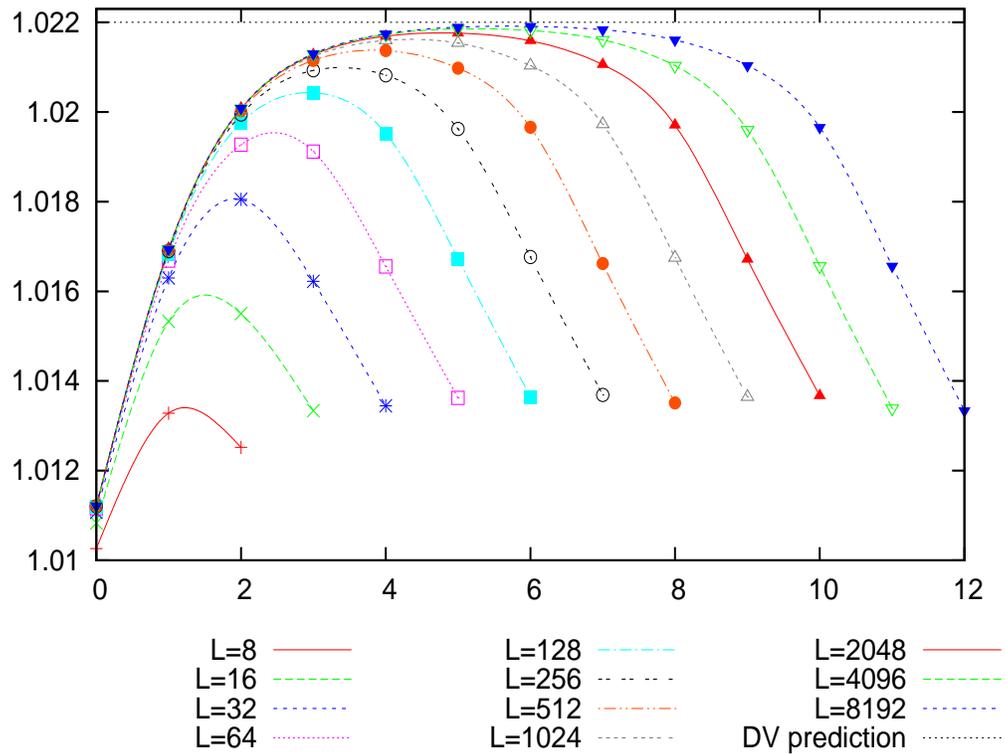}}
\end{center}
\caption{
Values of $R_{FK}$ vs $\log_2 \Delta $ for the bond percolation (or Potts model for $Q=1$), 
with $\Delta$ the linear size of the triangle employed to define the three-point correlation function 
(see definition in the text). The different curves correspond to interpolation between the points 
obtained for different linear sizes as shown in the caption. 
We also show DV prediction for that case: $R_{FK}= 1.022$ \cite{DV} as a dotted line.
}
\label{PQ1}
\end{figure}
We first show the result for the case $Q=1$ which corresponds to percolation. 
This case was already considered by  Ziff \textit{et al.} \cite{Ziff}. These authors considered the same correlation functions 
on the triangular lattice, the three points being the vertices of an equilateral triangle.
Our simulation is different since we consider a triangle with two edges of length $\Delta$ and one edge of length $\sqrt{2} \Delta$.
This difference must not affect the general result apart for finite size effects and indeed,
our findings, shown  in Fig.~\ref{PQ1}, are in excellent agreement with the ones in \cite{Ziff}.
In this figure, we plot the ratio $R_{FK}$ defined in (\ref{Rfk}) as a function of $\log_2 \Delta$ and for increasing values of linear size $L$. We also show DV prediction, $R_{FK}=1.0220$. 
A more quantitative check of the agreement is made in the following way. In Fig.~\ref{PQ1}, it is apparent that the curves have a maximum for $\Delta=\sqrt{(L/2)}$ which 
lies in the bulk asymptotic region $ 1 \ll \Delta \ll L$. We can compare the measured values of $R_{FK}(\sqrt{(L/2)})$ to a fit of the following form 
\beq
\label{fitR}
R_{FK}(\sqrt{(L/2)}) = R_0 + R_1 L^{-\omega}\; ,
\eeq
where $R_0$ is the constant to be compared to DV prediction. 
We obtain (keeping only the values for which $\sqrt{(L/2)}$ is a power of $2$) : 
$R_0 = 1.02197 (6)$ with a fit in the range $L=8-8192$, $R_0=1.02190 (3)$ with 
$L=32-8192$ and $R_0=1.02187 (4)$ with $L=128-8192$.  An  extrapolated value is 1.0218 (2) in very good agreement with DV prediction
\begin{equation}
 R_{FK}=\sqrt{2} C(\alpha_{\frac{3}{2},\frac{3}{2}},\alpha_{\frac{3}{2},\frac{3}{2}},\alpha_{\frac{3}{2},\frac{3}{2}})\sim1.0220
\end{equation}
\noindent 
{\bf \underline{Ising model, $Q=2$}}

\begin{figure}
\begin{center}
\epsfxsize=400pt\epsfysize=300pt{\epsffile{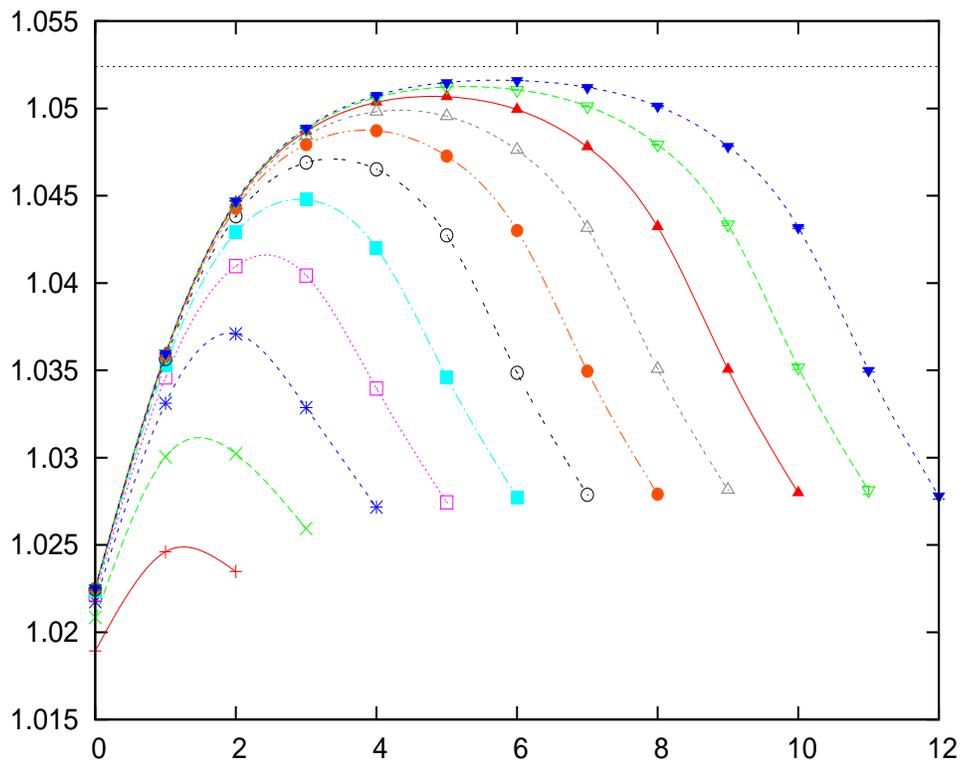}}
\end{center}
\caption{
Values of $R_{FK}$ vs $\log_2 \Delta $ for the Ising model ($Q=2$) FK clusters. DV prediction $R_{FK}=1.0524$ is shown as 
a dotted line. The colors for different sizes $L$ are the same as in Fig.~\ref{PQ1}.
}
\label{PQ2}
\end{figure}

\noindent 
In Fig.~\ref{PQ2}, we present our results for the Ising model. We show the numerical values for $R_{FK}(\Delta)$ as a function 
of $\log_2 \Delta $ as well as the DV prediction.
By performing the fit (\ref{fitR}), we obtain a value $R_0=1.0524 (2)$ in perfect agreement with the predicted value
\begin{equation}
\label{FK_ising}
R_{FK}=\sqrt{2} C(\alpha_{2,2},\alpha_{2,2},\alpha_{2,2})\sim 1.0524
\end{equation}

It is important here to observe that in the $c=1/2$ minimal model describing the Ising critical point, the correlation
\begin{equation}
\langle \Phi_{2,2}\Phi_{2,2}\Phi_{2,2}\rangle =\langle \Phi_{2,1}\Phi_{2,1}\Phi_{2,1}\rangle =0
\label{3spinIs}
\end{equation}
vanishes. This can be understood  from the vanishing of the 
three-spin correlation function by simple parity arguments, as it is evident in  (\ref{G3}) for $Q=2$.  In the CFT approach,
one can show that the correlation (\ref{3spinIs}) vanishes because it does not satisfy the fusion rules $\Phi_{2,1}\Phi_{2,1}=\Phi_{3,1}+\Phi_{1,1}$ imposed by the null vector condition. Indeed,
the spectrum of the minimal models is built by irreducible  Virasoro representation $\Phi_{n,m}$ whose null vector is assumed to decouple from the theory, i.e. 
the correlation functions containing a null vector have to vanish \cite{BPZ}.  As noticed in \cite{Za}, the time-like Liouville function (\ref{Zam_structure}) 
does not vanish automatically when fusion rules are not satisfied, the constant  $C(\alpha_{2,2},\alpha_{2,2},\alpha_{2,2})$ being an example. If one assumes 
the decoupling of the null vector from the theory, then the function  $C(\alpha_{n,m},\alpha_{r,s},\alpha_{p,q})$, for positive integers $n,m,p,q,r,s$
 has a meaning only when the fusion rules are satisfied. 
Otherwise, a clear interpretation of the values of the type $C(\alpha_{2,2},\alpha_{2,2},\alpha_{2,2})$  has remained previously unknown.  
It is then quite remarkable that  the three-point connectivity of the Ising FK clusters  provides such a natural interpretation. 
Note that this implies that, in the study of the percolative properties of the Ising model, the Virasoro representation can be reducible and the null vector states 
do not decouple from the theory. In certain respects,  this is not so surprising. Consider, for instance, the non-vanishing of $C(\alpha_1,\alpha_2,0)$, see (\ref{Zam_structure}), when the charges $\alpha_1$ and $\alpha_2$ are such that $\alpha_2\neq \alpha_1$ and $\alpha_2\neq \beta-\beta^{-1}-\alpha$. This implies the existence of a dimension $0$ primary which has a non-vanishing null vector at order $1$. This fact has to be understood by assuming the existence of an operator of dimension $0$ which is not the identity.  In the study of SLE processes, describing the evolution of boundary interfaces, such operators appear as bulk spectator operators. The presence of such operators  is crucial to determine the conformal invariant probabilities associated to SLE evolutions \cite{BaBe}. 
Finally, it has to be stressed that, at the Ising point, whose universality class is characterized by a $S_2$ symmetry, the three-point connectivity of Ising FK clusters `remembers'  the $S_Q$ symmetry of (\ref{mu2}), (\ref{mu3}), as the factor $\sqrt{2}$ in (\ref{FK_ising}) indicates.

\noindent 
{\bf \underline{$Q=3$ Potts model}}

\begin{figure}
\begin{center}
\epsfxsize=400pt\epsfysize=300pt{\epsffile{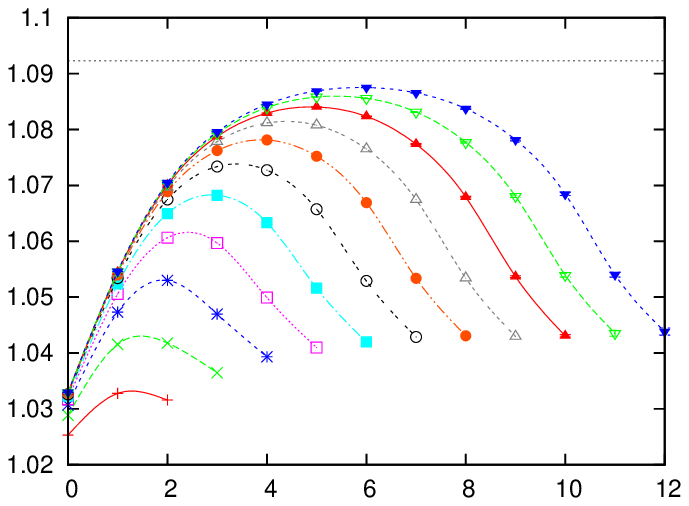}}
\end{center}
\caption{
Values of $R_{FK}$ vs $\log_2 \Delta$ for the $3$-state Potts model FK clusters. The dotted line corresponds to DV prediction 
 $R_{FK}=1.0923$. The colors for different sizes $L$ are the same as in Fig.~\ref{PQ1}.
}
\label{PQ3}
\end{figure}

\noindent 
As shown in the previous section, $Q=3$ is the only  case for which (\ref{Rfk}) can be derived with standard arguments. The numerical results, shown in Fig.~\ref{PQ3}, can 
be considered as a support of the validity of our numerical analysis.
Using the  fit (\ref{fitR}), one obtains the value $R_0 = 1.0925 (2)$ in excellent agreement with the exact result (\ref{FK_q3}), $R_{FK}=1.0923$.

\noindent 
{\bf \underline{$Q=4$ Potts model}}

\begin{figure}
\begin{center}
\epsfxsize=400pt\epsfysize=300pt{\epsffile{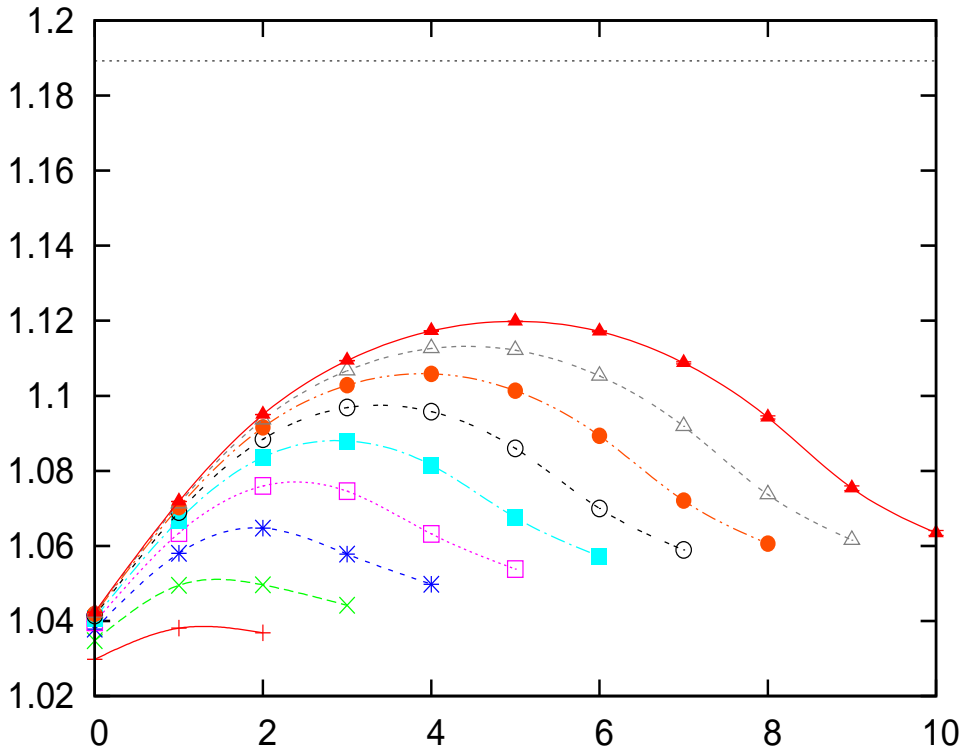}}
\end{center}
\caption{
Values of $R_{FK}$ vs $\log_2 \Delta$ for the $4$-state Potts model FK clusters. The dotted line corresponds to DV prediction  $R_{FK}=1.1892$. 
The colors for different sizes $L$ are the same as in Fig.~\ref{PQ1}.
\label{PQ4}
}
\end{figure}

\noindent 
%
%For the case of the $Q=4$ states Potts model, we expect the same result for both ratio, either the one 
%build from the FK clusters or the spin clusters. Indeed, in that case, the same universality class 
%is expected for both observables. 
In Fig.~\ref{PQ4}, we show our results for $Q=4$ as well as the DV prediction. 
Note that in this case, we have data only for sizes up to $L=2048$. The reason is that 
for $Q=4$, the autocorrelation time $\tau$ (corresponding to the number of cluster 
updates that we need to perform between two independent configurations) is very 
large. For $L=2048$, we determined $\tau \simeq 10 000$ for $Q=4$ while it was 
$\tau \simeq 15$ for $Q=2$ and $\tau \simeq 400$ for $Q=3$.
Nevertheless, even with data up to size $L=2048$, we see that the behavior is similar to what we obtained for 
other values of $Q$'s. The main difference is that in this case we observe much stronger finite size 
corrections. This is not unexpected since we know that the $Q=4$ states 
Potts model contains multiplicative logarithmic corrections. The origin of these corrections is well known \cite{NaSc,CaNaSc}.
They correspond to the merging of the critical and tricritical points of the $Q$-state Potts model for $Q=4$,
where the dilution field becomes marginal.  We also performed a fit to the form  (\ref{fitR}) 
for which we obtain $R_0 = 1.17 (2) $ which is compatible with the DV prediction $R_{FK} = 1.1892$.
Note that in this case the error on $R_0$ is very large. This is due to the fact that we have data only up to $L=2048$
but also to the fact that the correction term in (\ref{fitR}) is very small, \ie\ $\omega \simeq 0.16$. 
The smallness of the correction exponent can probably be related to the existence of logarithmic corrections
but we have not enough data points to check such terms. Thus we can not exclude the possibility  
that the agreement is just due to the fact that we have large errors in the fitting procedure.

\noindent 
{\bf \underline{Non Integer cases}}

\noindent 
Finally, we  present results for the $Q$-state Potts model when  $Q$ is not integer. This is also interesting because the $R_{FK}$ is expected to be given by
correlation functions of $\Phi_{(p+1)/2,(p+1)/2}$ operators with irrational scaling dimensions (i.e $p$ irrational). Studying the FK three-point connectivity for general $Q$, we are therefore probing correlations of CFTs which are beyond the logarithmic minimal models, known to play a role in the study of extended objects in critical systems.

\begin{figure}
\begin{center}
\epsfxsize=220pt\epsfysize=170pt{\epsffile{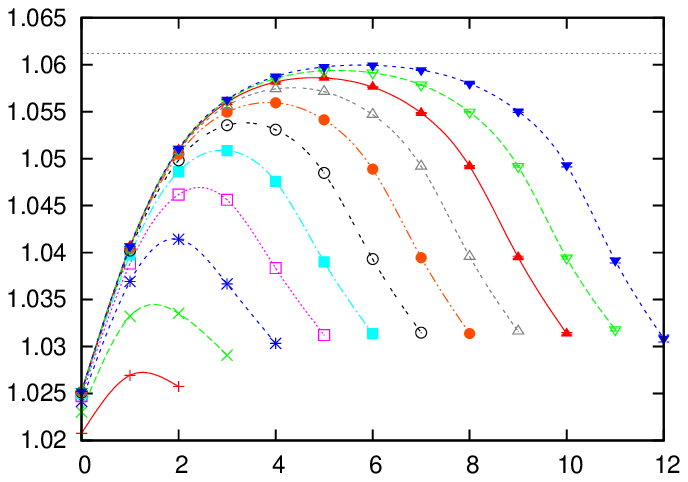}}
\epsfxsize=220pt\epsfysize=170pt{\epsffile{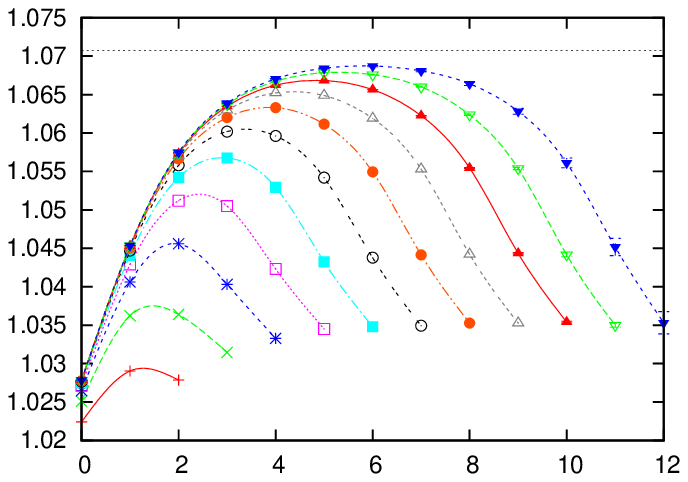}}\\
\epsfxsize=220pt\epsfysize=170pt{\epsffile{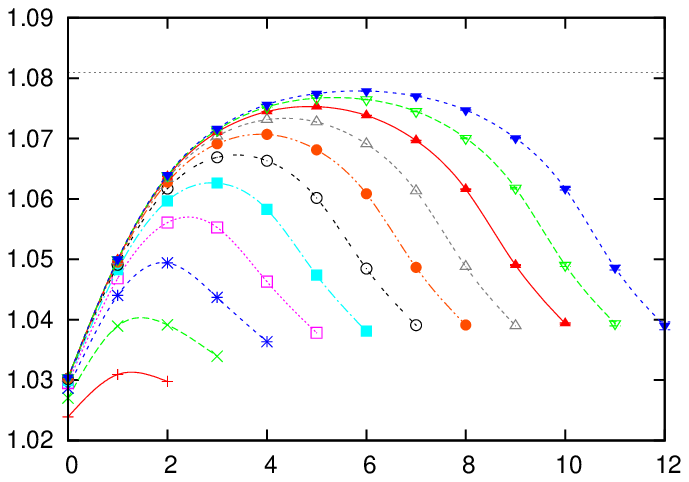}}
\end{center}
\caption{
$R_{FK}$ vs. $\log_2 \Delta$  for $Q=2.25$ in the upper left figure, $Q=2.5$ in the upper right figure and $Q=2.75$ 
in the lower figure. The colors for the $L$'s are the same as in Fig.~\ref{PQ1}.
\label{PQ2.5}
}
\end{figure}

%In that case we can consider only FK clusters. 
In Fig.~\ref{PQ2.5}, we present the results for $Q=2.25, 2.5$ and $Q=2.75$ corresponding respectively  
to $\; R_{FK}=1.0612, 1.0707, 1.0809$.

The results for the FK clusters for all the simulated values of Q's are reported in Tab.~\ref{Table1} and show an excellent 
 agreement with DV prediction. As we already mentioned, the case  $Q=4$ is peculiar since 
logarithmic corrections to the scaling are expected. Nevertheless, the agreement with the theoretical prediction remains quite good.

\begin{table}
\caption{Results for $R_{FK}$ for the FK clusters. The first line contains the results obtained from an extrapolation 
of the constant part $R_0$ in the fit to the form (\ref{fitR}) of our numerical data. The second line contains the prediction $R_{FK}$ of  \cite{DV}.}
\begin{center}
\begin{tabular}{ | l ||  c | c | c | c |  c |  c | c | }
\hline
  $Q$ & 1 & 2 & 2.25 & 2.5 & 2.75 & 3 & 4 \\ \hline 
  $R_0$ & 1.0218 (2)  & 1.0524 (2)  & 1.0613 (2) & 1.0706 (2) & 1.0811 (2) & 1.0925 (2) & 1.17 (2)  \\
  $R_{FK}$ & 1.022  & 1.0524   & 1.0612 & 1.0707 & 1.0809 &  1.0923 &  1.1892  \\
  \hline
\end{tabular}
\label{Table1}
\end{center}
\end{table}

\section{Conclusion}

In this paper we considered the three-point connectivity of FK clusters in the Q-state Potts model. 
On the theoretical side, we presented a very detailed analysis of the derivation of the time-like Liouville correlator within the Coulomb gas approach. In particular we 
showed that, on the basis of very general assumptions, such as conformal invariance and absence of degeneracy in the spectrum of scaling 
dimensions, the time-like Liouville correlator has to be considered as the only analytic continuation of minimal model structure constants. 
We then checked the relation (\ref{Rfk}) between the constant $R_{FK}$  and the time-like Liouville correlator. 
Numerical simulations were performed for integer and non integer values of $Q$, giving a striking support to this conjecture. 

We stressed that, together with the percolation case $Q=1$, already tested numerically in \cite{Ziff}, the confirmation of the theoretical prediction for the Ising case $Q=2$ is particularly interesting. Indeed, it was observed in \cite{Za} that the structure constants (\ref{Zam_structure}) do not always vanish when specialized to minimal cases for which the minimal model OPE prescribes a zero. The finite numbers that (\ref{Zam_structure}) yields instead were defined `mysterious' in \cite{Za}. One such number arises at the Ising central charge 1/2 for three fields with the conformal dimension 1/16 of the Ising spin field. Within the minimal CFT description of the Ising model this vertex is absent, as required by the spin reversal symmetry of the model, which is implemented in (\ref{G3}) by the factor $Q-2$. The argument of \cite{DV}, however, relates Zamolodchikov's formula to the connectivity (\ref{mu3}), which does not vanish. This illustrates how the `mysterious' numbers of \cite{Za} may acquire a 
physical interpretation in relation to observables, like cluster connectivities, implementing a non-minimal realization of the conformal symmetry.

It would be interesting to test this emerging scenario by studying  other geometric  observables such as four-point connectivities.  This in 
particular would shed light on the self-consistency of time-like Liouville CFT.
Finally, an important issue that we will consider in \cite{DPSV2} is that of the three-point connectivity for spin clusters.

%\begin{equation}
%\label{spin_clu}
% R_{S} \stackrel{?}{\propto}
% C(\alpha_{\frac{p}{2},\frac{p}{2}},\alpha_{\frac{p}{2},\frac{p}{2}},\alpha_{\frac{p}{2},\frac{p}{2}}).
%\end{equation}

\vspace{1cm}
\noindent
{\bf ACKNOWLEDGMENTS:}
We thank Vladimir Dotsenko, Jesper Jacobsen, Sylvain Ribault and Romain Vasseur for very useful discussions. R.S.  acknowledges support by ANR grant
2011-BS04-013-01 WALKMAT.

\end{document}